\documentclass[10pt, conference, twocolumn]{IEEEtran}
\usepackage{amsfonts}
\usepackage{amssymb}
\usepackage{graphicx}
\usepackage{amsmath}
\usepackage{multirow}
\usepackage{mathtools}
\usepackage{graphicx}
\usepackage{subcaption}
\usepackage{cite}
\usepackage{color}
\usepackage{xcolor}
\usepackage{soul}
\usepackage{makecell}
\usepackage{booktabs}
\usepackage{microtype}
\usepackage{textcomp}
\usepackage{siunitx}
\usepackage{tabu}
\usepackage{float}
\usepackage{tabularx}
\usepackage{listings}
\usepackage{algorithm}
\usepackage{algpseudocode}
\newcolumntype{C}{ >{\centering\arraybackslash}X}
\usepackage{bm}
\usepackage{threeparttable}
\usepackage{gensymb}
\def\degree{${}^{\circ}$}

\definecolor{Trypan_blue}{RGB}{58, 12, 163}
\definecolor{grey}{RGB}{245,245,245}
\definecolor{blue}{RGB}{43,159,236}
\definecolor{green}{RGB}{19,158,33}

\usepackage[version=3]{acro}
\acsetup{
	use-id-as-short=true,
	make-links=false,
	link-only-first=true,
}
\makeatletter

\newcommand{\Rmnum}[1]{\expandafter\@slowromancap\romannumeral #1@}

\makeatother
\DeclareAcronym{AI}{
	long={artificial intelligence},
}
\DeclareAcronym{ISLs}{
	long={inter-satellite links},
}
\DeclareAcronym{MARL}{
	long={multi-agent reinforcement learning},
}
\DeclareAcronym{DRL}{
	long={deep reinforcement learning},
}
\DeclareAcronym{LLM}{
	long={large language model},
}
\DeclareAcronym{MEC}{
    long={mobile edge computing},
}
\DeclareAcronym{GPU}{
	long={graphics processing unit},
}
\DeclareAcronym{MAE}{
	long={mean absolute error},
}
\DeclareAcronym{MSE}{
	long={mean square error},
}
\DeclareAcronym{NMSE}{
	long={normalized mean square error},
}
\DeclareAcronym{DL}{
	long={deep learning},
}
\DeclareAcronym{LEO}{
    long={low Earth orbit},
}
\DeclareAcronym{SWaP}{
	long={size, weight and power},
}
\DeclareAcronym{CPU}{
	long={central processing unit},
}
\DeclareAcronym{MADQN}{
    long={Multi-agent deep Q-network},
}
\DeclareAcronym{IDQN}{
    long={Independent deep Q-network},
}
\DeclareAcronym{MAPPO}{
    long={Multi-agent proximal policy optimization},
}
\DeclareAcronym{PPO}{
    long={proximal policy optimization},
}
\DeclareAcronym{ORCA}{
    long={Onboard Reflective Collaborative Agents},
}
\DeclareAcronym{E2E}{
    long={end-to-end},
}
\DeclareAcronym{6G}{
	long={the sixth-generation},
}

\graphicspath{{Figures/}}

\begin{document}
	 \title{ORCA: Onboard Reflective Collaborative Agents for Reliable Task Offloading in LEO Satellite Networks}
     
\author{
	\IEEEauthorblockN{
		Yunxin Geng\IEEEauthorrefmark{1}\IEEEauthorrefmark{2},
        Shengxiang Hao\IEEEauthorrefmark{1}\IEEEauthorrefmark{2},
        Yichen Wu\IEEEauthorrefmark{1}\IEEEauthorrefmark{2},
        Haotian Liu\IEEEauthorrefmark{1}\IEEEauthorrefmark{2},\\
        Weiwei Jiang\IEEEauthorrefmark{1}\IEEEauthorrefmark{2},
        Lu Cao\IEEEauthorrefmark{1}\IEEEauthorrefmark{3},
        Han Meng\IEEEauthorrefmark{4},
        Yuxuan Pan\IEEEauthorrefmark{5},
		and Zhiqing Wei\IEEEauthorrefmark{1}\IEEEauthorrefmark{2}} 
    \IEEEauthorblockA{\IEEEauthorrefmark{1}Key Laboratory of Universal Wireless Communication, Ministry of Education, Beijing, China}
    \IEEEauthorblockA{\IEEEauthorrefmark{2}Beijing University of Posts and Telecommunications, Beijing, China}

    \IEEEauthorblockA{\IEEEauthorrefmark{3}‌Xidian University, Xi'an, China}
    \IEEEauthorblockA{\IEEEauthorrefmark{4}Institute of Network and IT Technology, China Mobile Research Institute, Beijing, China}
    \IEEEauthorblockA{\IEEEauthorrefmark{5}The 7th Research Institute of China Electronics Technology Group Corporation, Guangzhou, China}
Email: \{gengyx, haoshengxiang, wuyichen, haotian\_liu,  jww, panyx, weizhiqing\}@bupt.edu.cn, \\
25173121481@stu.xidian.edu.cn,
menghanyjy@chinamobile.com
}



\maketitle

\begin{abstract}
Low Earth orbit (LEO) satellites increasingly perform time-critical onboard processing, yet heterogeneous computing capabilities, rapidly changing inter-satellite links (ISLs) and uncertain neighbor behavior make decentralized task offloading difficult under partial observations.
This paper presents ORCA, an architecture of Onboard Reflective Collaborative Agents for autonomous onboard task offloading. ORCA organizes offloading decisions into a four-Agent \emph{Perceive}$\rightarrow$\emph{Reason}$\rightarrow$ \emph{Act}$\rightarrow$\emph{Learn} loop. It aims to improve efficiency and reliability in dynamic, heterogeneous, and unreliable LEO satellite networks.
The simulation results of a heterogeneous 60-satellite Walker-delta constellation demonstrates that
ORCA delivers competitive system efficiency in cooperative settings while achieving superior task-success robustness and offloading reliability under previously unseen malicious node behavior,
indicating that outcome-grounded reflection can improve the reliability of decentralized onboard offloading without centralized runtime coordination.
\end{abstract}


\section{Introduction}
\begin{figure*}[!t]
    \centering
    \includegraphics[width=0.99\textwidth]{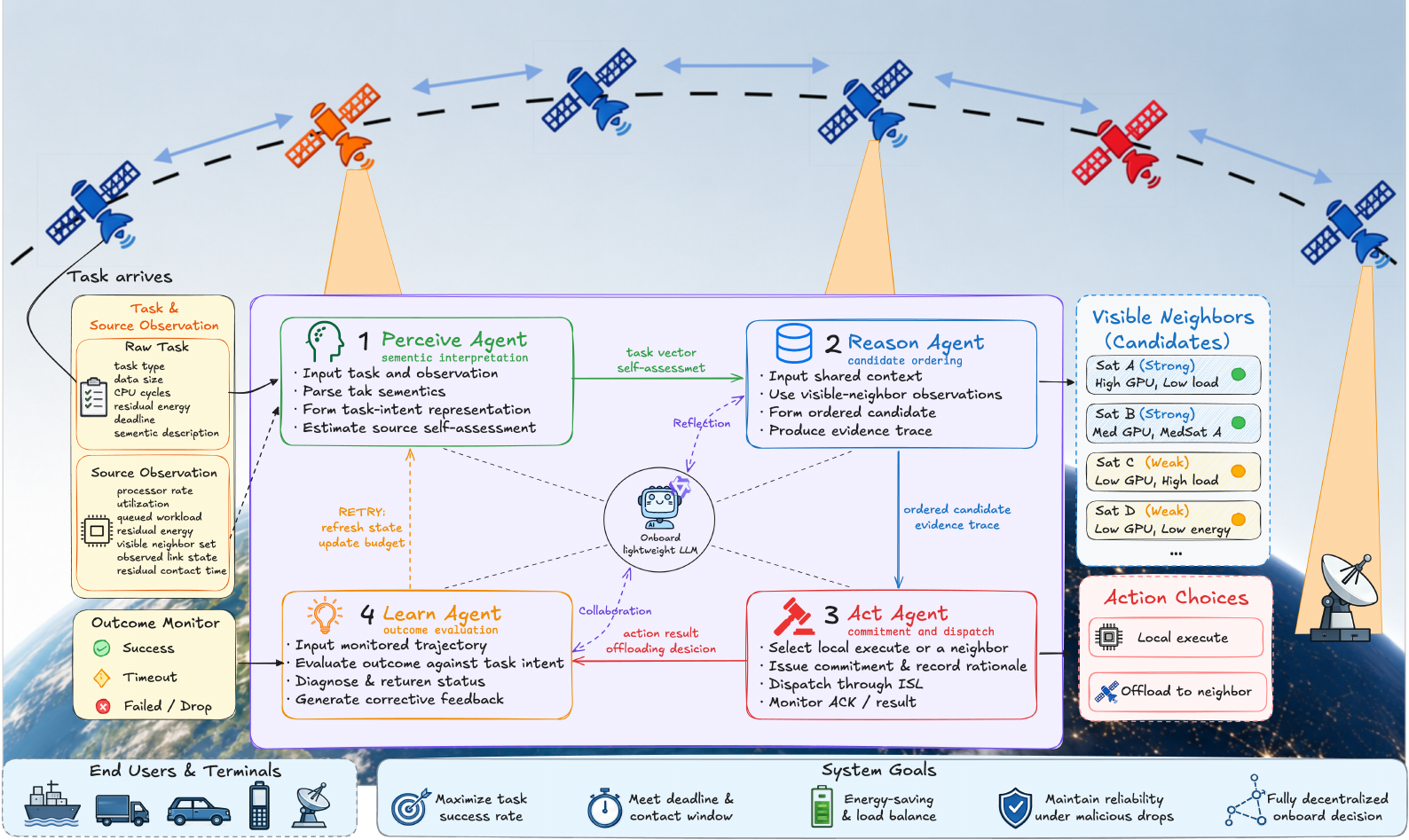}
    \caption{Illustration of the proposed four-Agent \emph{Perceive}$\rightarrow$\emph{Reason}$\rightarrow$\emph{Act}$\rightarrow$\emph{Learn} ORCA architecture. 
    }
    \label{fig:framework}
\end{figure*}

Advances in remote sensing and onboard \ac{AI} are turning satellites into in-orbit computing platforms~\cite{guo2025enabling,dong2026fluidspatiotemporalstochasticgeometryinformation}.
Yet limited \ac{SWaP} constrains local execution, while ground-based offloading depends on intermittent contact.
For example, Sentinel-1 supports a $520~\mathrm{Mb/s}$ X-band data downlink and a $64~\mathrm{kb/s}$ S-band command uplink, whereas the EDRS LEO-to-GEO optical link supports up to $1.8~\mathrm{Gb/s}$.
These mission-specific rates highlight the potential of high-capacity \ac{ISLs} for sharing time-critical workloads without waiting for ground access.

Nevertheless, such offloading faces partial and rapidly changing neighbor observations, heterogeneous computing loads, and finite contact windows.
These factors complicate deadline-aware task commitment; moreover, a selected neighbor may accept a task without completing it.

Existing studies formulate satellite offloading as online optimization or learn policies through \ac{DRL} and \ac{MARL}~\cite{yu2022surprising,chen2025ms_m2atd3}.
\ac{DRL}-based methods learn sequential policies instead for user association, computing-satellite selection and multidimensional resource allocation~\cite{zhang2024dynamic}. 
As the network scale and decision coupling increase, \ac{MARL} distributes decision-making among multiple users or satellite nodes. 
In \ac{LEO} satellite edge computing, hierarchical \ac{MARL} separates computation offloading and resource allocation across different control levels~\cite{lai2024joint}. Another approach, primary-secondary \ac{MARL}, coordinates a primary agent and multiple local agents for energy-efficient task offloading~\cite{chen2025ms_m2atd3}.
However, these approaches still learn policies within predefined Markov games, where agent roles, observations, actions, rewards and interaction structures are fixed before training. 
More recently, Agentic~\ac{AI} has extended autonomous network management through goal interpretation, task decomposition, multi-step planning, memory, and execution feedback.
\ac{LLM} based Agents have been investigated for perception and alignment in \ac{6G} network~\cite{xu2024llm}, decentralized network optimization~\cite{lee2025aidriven} and intent-driven infrastructure orchestration~\cite{brodimas2025internet}. 
Nevertheless, existing studies mainly focus on global network management and service orchestration. Onboard \ac{LEO} task offloading remains largely unexplored under partial observations, heterogeneous capabilities, short contact windows, and uncertain collaboration outcomes.


Reliable \ac{LEO} offloading thus requires task interpretation, evidence-based commitment, outcome monitoring, and timely correction.
To this end, we propose \ac{ORCA}, a four-Agent onboard \emph{Perceive}$\rightarrow$\emph{Reason}$\rightarrow$\emph{Act}$\rightarrow$\emph{Learn} loop that goes beyond a single policy invocation.
The \emph{Learn} Agent returns outcome-grounded feedback for a bounded corrective pass.

The main contributions of this paper are as follows:
\begin{itemize}
\item \textbf{Four-Agent autonomous offloading architecture:} 
\ac{ORCA} decomposes task offloading into semantic perception, contextual candidate reasoning, physical commitment and reflective outcome evaluation. 
This closed loop operates from onboard observations without centralized runtime coordination or evaluation-time gradient updates.
\item \textbf{Reliability under an unseen neighbor-failure process:} 
A 60-satellite ground simulation is conducted to compare \ac{ORCA} with classical baselines in cooperative and unreliable settings, 
indicating that \ac{ORCA} maintains the strongest task-success robustness under the unseen malicious setting while preserving competitive system efficiency.
\end{itemize}

The remainder of this paper is organized as follows.
Section~\ref{sec:problem_formulation} introduces satellite computation offloading, and the proposed \ac{ORCA} framework is described  in Section~\ref{sec:system_model}.
The ground simulation is  conducted in Section~\ref{sec:experiment}  as well as the analysis of the results. 
Section~\ref{sec:conclusion} concludes the paper.

\section{Problem Formulation}
\label{sec:problem_formulation}
Consider a constellation $\mathcal S=\{1,\ldots,N\}$ with directed ISL graph $\mathcal G_t=(\mathcal S,\mathcal E_t)$ at time $t$. 
Let $\tau_k=(d_k,c_k,\Delta_k,\boldsymbol\eta_k)$ be the $k^{\text{th}}$ task arrives at source satellite $i$ at time $t$, 
where $d_k$ is the data size, $c_k$ is the required number of \ac{CPU} cycles, $\Delta_k$ is the relative deadline and $\boldsymbol\eta_k$ is the semantic task description. 
The source satellite observes
\begin{equation}
\mathbf o_{i_k,t}=\bigl(\tau_k,\mathbf s_{i,t},\mathcal N_{i,t},
\mathbf g_{i,t}\bigr),
\label{eq:observation}
\end{equation}
where $\mathbf s_{i,t}=(f_{i,t},\rho_{i,t},Q_{i,t},E_{i,t})$ contains the processor rate, utilization, queued workload and residual energy. 
The visible-neighbor set is defined as $\mathcal N_{i,t}=\{j:(i,j)\in\mathcal E_t\}$, and $\mathbf g_{i,t}$ contains the observed link state and residual contact time. 
Thus, no global constellation state is required at the source.

Let $t_k^{\mathrm c}$ denote the time when the terminal commitment for the $k^{\text{th}}$ task is issued. A corrective round may precede this commitment. Let $b_k^{\mathrm c}$ denote the service budget remaining at that time.
Variable $x_{k,j}\in\{0,1\}$ indicates that executor $j\in\{i\}\cup\mathcal N_{i,t_k^{\mathrm c}}$ receives the commitment. 
The deadline-accounted service time $D_{k,j}$ contains queueing and execution, together with transmission and the commitment handshake for a remote executor. 
The reported system delay $L_k$ also includes the configured Agent deliberation overhead and any preceding failed attempts. 
The task-success indicator is
\begin{equation}
    I_k=\sum_jx_{k,j}\mathbf 1\{D_{k,j}\leq b_k^{\mathrm c}\text{ and task $k$ completes}\}.
    \label{eq:indicator}
\end{equation}
Here $I_k\in\{0,1\}$ denotes completion within the configured service deadline, therefore the task utility can be calculated by
\begin{equation}
    U_k=w_sI_k-w_l\frac{\bar L_k}{\Delta_k}
    -w_e\frac{\varepsilon_k}{\varepsilon_{\mathrm{ref}}}.
    \label{eq:utility}
\end{equation}
The capped system delay is $\bar L_k=\min(L_k,\Delta_k)$ for a completed task and $\bar L_k=\Delta_k$ otherwise, preventing an early failure from receiving a small delay penalty. 
The cumulative task energy $\varepsilon_k$ includes deliberation, transmission and execution energy incurred over all attempted rounds. 
The nonnegative weights $w_s$, $w_l$, and $w_e$ encode mission priorities, while $\varepsilon_{\mathrm{ref}}$ normalizes energy.

For tasks $\mathcal K$, autonomous policies $\boldsymbol\pi=\{\pi_i:i\in\mathcal S\}$ seek
\begin{align}
\max_{\boldsymbol\pi}\quad &\mathbb E_{\boldsymbol\pi}
\left[|\mathcal K|^{-1}\sum_{k\in\mathcal K}U_k\right] \label{eq:objective}\\
\mathrm{s.t.}\quad
&\sum_{j\in\{i\}\cup\mathcal N_{i,t_k^{\mathrm c}}}x_{k,j}\leq1, \label{eq:unique}\\
&x_{k,j}\widehat D_{k,j}\leq x_{k,j}b_k^{\mathrm c},
\quad \forall j, \label{eq:deadline}\\
&x_{k,j}(\widehat T_{k,ij}^{\mathrm{hs}}+\widehat T_{k,ij}^{\mathrm{tx}}\nonumber\\
&\quad+\widehat T_{k,j}^{\mathrm q}+\widehat T_{k,j}^{\mathrm{exe}})
\leq x_{k,j}V_{ij,t_k^{\mathrm c}},\quad \forall j\neq i. \label{eq:visibility}
\end{align}
Eq.~\eqref{eq:unique} permits at most one terminal executor, with equality to zero representing deferral when no feasible commitment remains. 
$\widehat D_{k,j}$ in Eq.~\eqref{eq:deadline} is the service time estimated at $t_k^{\mathrm c}$, whereas $D_{k,j}$ in Eq.~\eqref{eq:indicator} is the realized value.
The residual budget $b_k^{\mathrm c}$ equals $\Delta_k$ for the first commitment and is reduced by the elapsed time of an unsuccessful round before a corrective commitment. 
$\widehat T_{k,ij}^{\mathrm{hs}}$, $\widehat T_{k,ij}^{\mathrm{tx}}$, $\widehat T_{k,j}^{\mathrm q}$ and $\widehat T_{k,j}^{\mathrm{exe}}$ in Eq.~\eqref{eq:visibility} are estimated handshake, transmission, queueing and execution times, and $V_{ij,t_k^{\mathrm c}}$ is the residual contact duration. 
The remote execution and its acknowledgment must remain within this interval. 
\ac{ORCA} constructs feasible task-level commitments online rather than through a centralized controller.

\section{System Model}
\label{sec:system_model}
As illustrated in Fig.~\ref{fig:framework}, \ac{ORCA} realizes policy $\pi_i$ through an onboard Agentic \emph{Perceive-Reason-Act-Learn} loop.

\subsection{Agentic Perceive--Reason--Act--Learn Loop}
As summarized in Algorithm~\ref{alg:orca}, upon receiving the task, the source satellite $i_k$ records local state and \ac{ORCA} loop is instantiated based on the onboard lightweight \ac{LLM}, whose parameters are $\theta_{i}$. 
Let $\mathcal P$ denote the \emph{Perceive} Agent, we have
\begin{equation}
(\mathbf v_k,a_k)=\mathcal P_{\theta_{i}}(\tau_k,\mathbf o_{i_k,t_k^1}),
\label{eq:perceive}
\end{equation}
where $\mathbf v_k$ and $a_k$ are the task-intent representation and source self-assessment, respectively.
At the $r^\text{th}$ deliberation round, the shared context is
\begin{equation}
\mathbf h_k^r=\bigl(\mathbf o_{i,t_k^r},\mathbf v_k,a_k,
\mathcal Z_{i,t_k^r},\boldsymbol\delta_k^{r-1},b_k^r\bigr),
\label{eq:context}
\end{equation}
where $\mathbf o_{i,t_k^r}$ is the latest locally available state,  $\mathcal Z_{i,t_k^r}$ is a bounded record of recent role messages and observed outcomes, 
and vector $\boldsymbol\delta_k^{r-1}$ carries the preceding diagnosis.
The record supplies the interaction context needed for correction and is not globally shared.

The \emph{Perceive} Agent first translates the raw task and current source state into a common intent frame. The \emph{Reason} Agent $\mathcal R$ then combines this frame with visible-neighbor observations, local interaction context, and any prior diagnosis. It produces an ordered candidate list $C_k^r$ and its evidence trace $\boldsymbol\psi_k^r$:
\begin{equation}
    (C_k^r,\boldsymbol\psi_k^r)=\mathcal R_{\theta_{i}}(\mathbf h_k^r).
    \label{eq:reason}
\end{equation}
Then, \emph{Act} Agent $\mathcal A$ selects local execution or a visible neighbor and emits the corresponding commitment $u_k^r$ and rationale $z_k^r$:
\begin{equation}
    (u_k^r,z_k^r)=\mathcal A_{\theta_{i_k}}(\mathbf h_k^r,C_k^r,\boldsymbol\psi_k^r).
    \label{eq:act}
\end{equation}
After dispatch, the environment returns an execution outcome rather than an Agent-generated estimate.
\emph{Learn} Agent $\mathcal L$ then evaluates whether the observed outcome satisfies the interpreted task intent and identifies why another decision may be warranted.
It returns status $\chi_k^r\in\{\textsc{accept},\textsc{retry},\textsc{terminate}\}$ and corrective feedback $\boldsymbol\delta_k^r$:
\begin{equation}
    (\chi_k^r,\boldsymbol\delta_k^r)=\mathcal L_{\theta_{i_k}}(\mathcal T_k^r). 
    \label{eq:learn}
\end{equation}
Status \textsc{accept} closes an outcome that \emph{Learn} deems final, \textsc{retry} requests another decision if time remains, and \textsc{terminate} stops an infeasible or exhausted task.
Agents use distinct role prompts and fixed schemas rather than separately trained control policies.
\begin{algorithm}[t]
\caption{ORCA reflective four-Agent loop}
\label{alg:orca}
\scriptsize
\begin{algorithmic}[1]
\State \textbf{Initialize:} source $i_k$, round $r\gets1$, time $t_k^1\gets t_k$, budget $b_k^1\gets\Delta_k$, feedback $\boldsymbol\delta_k^0\gets\varnothing$, and maximum rounds $R_{\max}$.
\State Observe $\mathbf o_{i_k,t_k^1}$ and obtain $(\mathbf v_k,a_k)\gets\Call{Perceive}{\tau_k,\mathbf o_{i,t_k^1}}$.
\While{$r\leq R_{\max}$ and $b_k^r>0$}
  \State Form shared context $\mathbf h_k^r$ using Eq.~\eqref{eq:context}.
  \State $(C_k^r,\boldsymbol\psi_k^r)\gets\Call{Reason}{\mathbf h_k^r}$.
  \State $(u_k^r,z_k^r)\gets\Call{Act}{\mathbf h_k^r,C_k^r,\boldsymbol\psi_k^r}$.
  \State Dispatch $u_k^r$ and monitor $y_k^r$, $\Delta t_k^r$, $\Delta\varepsilon_k^r$, and $\mathbf o_{i,t_k^{r+1}}$.
  \State Build trajectory $\mathcal T_k^r$ using Eq.~\eqref{eq:trajectory}.
  \State $(\chi_k^r,\boldsymbol\delta_k^r)\gets\Call{Learn}{\mathcal T_k^r}$.
  \State Update budget, time, and transcript using Eq.~\eqref{eq:refresh}.
  \If{$\chi_k^r=\textsc{accept}$}
    \State \Return commitment $u_k^r$ and outcome $y_k^r$.
  \ElsIf{$\chi_k^r=\textsc{terminate}$}
    \State \Return terminal outcome $y_k^r$.
  \EndIf
  \State $r\gets r+1$.
\EndWhile
\State \Return the last monitored outcome with termination reason.
\end{algorithmic}
\end{algorithm}

\subsection{Monitored trajectory and reflective correction}
Execution of $u_k^r$ produces outcome $y_k^r$, elapsed time $\Delta t_k^r$, consumed energy $\Delta\varepsilon_k^r$ and post-action state $\mathbf o_{i,t_k^{r+1}}$. 
Here, $y_k^r$ records the completion status, executor, and any observed failure cause. 
\ac{ORCA} records the complete round as
\begin{equation}
\mathcal T_k^r=\bigl(\mathbf h_k^r,C_k^r,\boldsymbol\psi_k^r,u_k^r,z_k^r,
y_k^r,\Delta t_k^r,\Delta\varepsilon_k^r,\mathbf o_{i_k,t_k^{r+1}}\bigr),
\label{eq:trajectory}
\end{equation}
which is the monitored state, reasoning, action, and outcome trajectory.
Its entries cover the inputs and outputs of all four Agents.
The post-action state prevents evaluation from depending only on the rationale that preceded execution.

After \emph{Learn} evaluates a round, \ac{ORCA} updates time and local context according to
\begin{align}
b_k^{r+1}&=\max(0,b_k^r-\Delta t_k^r),\nonumber\\
t_k^{r+1}&=t_k^r+\Delta t_k^r,\nonumber\\
\mathcal Z_{i,t_k^{r+1}}
&=\operatorname{Trim}\!\left(\mathcal Z_{i,t_k^r}
\oplus(\mathcal T_k^r,\boldsymbol\delta_k^r);B_{\mathrm{ctx}}\right),
\label{eq:refresh}
\end{align}
where 
$\oplus$ appends the monitored trajectory and diagnosis.
The operator $\operatorname{Trim}(\cdot;B_{\mathrm{ctx}})$ enforces the local context-window budget $B_{\mathrm{ctx}}$.
If the token count exceeds this limit, it removes the oldest completed interaction records while retaining information needed for the next reflective pass.
The loop stops when \emph{Learn} accepts or terminates, the deadline budget reaches zero, or the round reaches $R_{\max}$. 
Reflection is therefore bounded by physical time and the configured interaction budget.
A retry carries \emph{Learn}'s diagnosis, residual deadline, post-action state, and bounded interaction record into the next \emph{Reason}-to-\emph{Act} pass.
The task intent remains fixed, while the operational context reflects the preceding outcome.

\section{Simulation Results and Analysis}
\label{sec:experiment}
This section firstly introduces the simulation setup and baseline methods for comparison, after which the performance of the proposed \ac{ORCA} is discussed.

\subsection{Simulation setup}
The ground simulation emulates onboard computing, ISLs and orbital motion for a 60-satellite Walker-delta constellation.
Table~\ref{tab:parameters} summarizes the constellation, workload, scenario and
\ac{ORCA} deployment parameters used in the evaluation.
Weak and strong satellites run the same four-agent protocol with different onboard INT4-quantized \ac{LLM}s and compute profiles.

The workload comprises image processing, signal demodulation, data compression and model inference,
with input size ranges from $1\mathord{\sim}5~\mathrm{MB}$, $0.5\mathord{\sim}2~\mathrm{MB}$, $1\mathord{\sim}5~\mathrm{MB}$ and $0.5\mathord{\sim}3~\mathrm{MB}$.
Their compute-demand settings are $5\pm2$, $2\pm0.8$, $3\pm1$ and $5\pm2~\mathrm{Gcycle}$, 
and their mean deadlines are $0.6~\mathrm{s}$, $0.3~\mathrm{s}$, $1.2~\mathrm{s}$ and $0.9~\mathrm{s}$.
The listed Agent delay and energy costs, including configured deliberation overhead, are configured simulation values and are averaged over all generated tasks.

\begin{table}[t]
\vspace{0.05in}
\centering
\caption{Experimental Parameters}
\label{tab:parameters}
\begin{tabular}{c|c}
\toprule
\textbf{Parameter} & \textbf{Value}\\
\midrule
\multicolumn{2}{c}{\textbf{Constellation and satellite parameters}}\\
\midrule
Number of satellites/orbital planes & 60 / 6\\
\midrule
LEO altitude/inclination & $600~\text{km}$ / $53\degree$\\
\midrule
ISL payload rate/range & $500~\mathrm{MB/s}$ / $5000~\mathrm{km}$\\
\midrule
ISL handshake latency/energy & 5 ms / 0.05 J\\
\midrule
Number of weak/strong satellites & 36 / 24\\
\midrule
Compute rate of weak/strong satellites & $4 / 12~\mathrm{Gcycle/s}$\\
\midrule
Compute power of weak/strong satellites & $15 / 25~\mathrm{W}$\\
\midrule
\multicolumn{2}{c}{\textbf{Task and scenario parameters}}\\
\midrule
Task-arrival rate & $60~\mathrm{tasks/min}$\\
\midrule
Burst workload & 5 tasks within $5~\mathrm{s}$\\
\midrule
Unreliable nodes/task-drop probability & 15 (25\%) / 0.9 \\
\midrule
\multicolumn{2}{c}{\textbf{ORCA framework and deliberation parameters}}\\
\midrule
Onboard \ac{LLM} on weak/strong satellites & Qwen3-8B/Qwen3-14B\\
\midrule
Maximum executed rounds & 2\\
\midrule
Full-role delay on weak/strong satellites & 0.50 s / 0.15 s\\
\midrule
Full-role energy on weak/strong satellites & 15 J / 5 J\\
\midrule
Context-reuse delay on weak/strong satellites & 0.08s / 0.03 s\\
\midrule
Context-reuse energy on weak/strong satellites & 2.0 J / 0.7 J \\
\bottomrule
\end{tabular}
\end{table}

To clearly demonstrate the effectiveness of the proposed \ac{ORCA}, some classic baseline methods are implemented for comparison.
\begin{itemize}
    \item \textbf{Random:} A visible neighbor is selected randomly, and local execution is used only when no neighbor satellite is visible.
    \item \textbf{\ac{IDQN}}~\cite{tampuu2017multiagent}\textbf{:} Each satellite maintains a separate two-layer Q-network with 64 units per layer and a 5000-transition replay buffer. 
    The learning rate is \num{1.0d-3}, the exploration rate $\epsilon$ decreases from 1.0 to 0.05 with a 0.99 decay factor, and the target network is updated every 100 steps.
    \item \textbf{\ac{MADQN}}~\cite{gupta2017cooperative}\textbf{:} The cooperative deep-Q baseline follows centralized cooperative experience sharing with decentralized action selection. 
    Satellites share a two-layer, 64-unit Q-network and a 50000-transition replay buffer.
    The learning rate, exploration schedule and target-update interval are the same with \ac{IDQN}.
    \item \textbf{\ac{MAPPO}}~\cite{yu2022surprising}\textbf{:} The shared stochastic actor and candidate-aware centralized critic use clipped \ac{PPO}. 
    The actor and critic learning rates are \num{3.0d-4} and \num{1.0d-3}, respectively, with a clipping ratio of 0.2 and a batch size of 256.
\end{itemize}
Agents of \ac{ORCA} retain only the bounded interaction record defined in Section~\ref{sec:system_model} and context reuse changes the configured deliberation cost but not the Agent interfaces or monitored commitment.
A cooperative scenario and a malicious one are evaluated.
In the malicious scenario, 25\% of satellites are randomly selected as unreliable and independently drop a task with a probability of 0.9 when sourcing it or after accepting a remote commitment.
All methods are run multiple times independently under five identical random seeds, with the averaged results and standard deviations recorded.

\subsection{Results and Analysis}
\subsubsection{Task success and zero-shot robustness}
\begin{figure}[tb]
    \centering
    \includegraphics[width=0.48\textwidth]{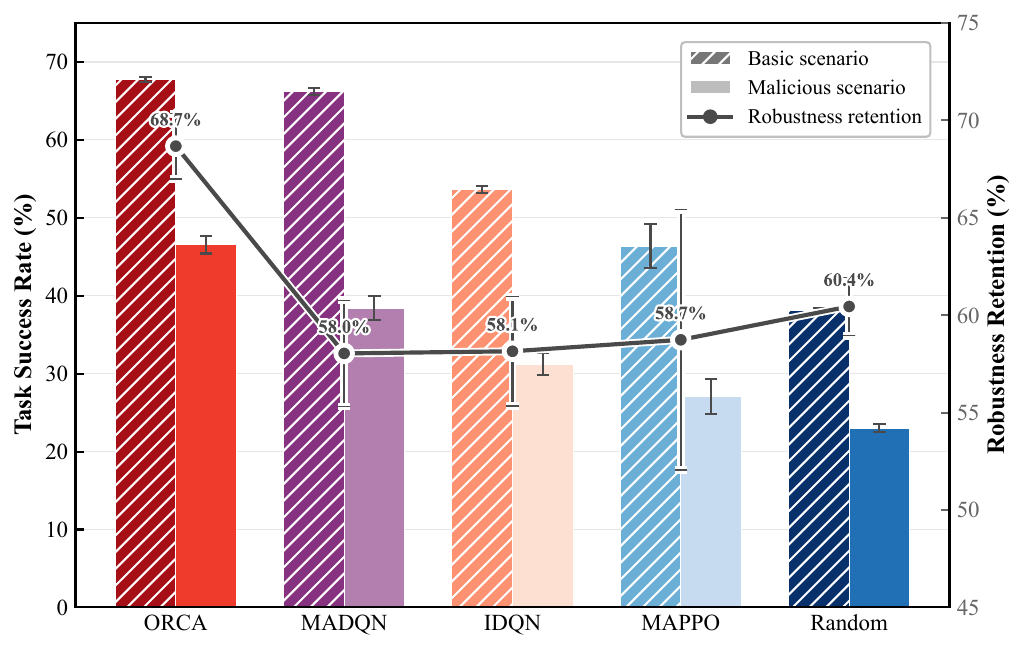}
    \caption{Comparison of task success rate in cooperate and malicious scenarios and the robustness-retention ratio across different methods.}
    \label{fig:task_success}
\end{figure}

\
\newline
\indent 
The task completion capability and robustness of \ac{ORCA} are first evaluated under cooperative and unseen malicious scenario.
As shown in Fig.~\ref{fig:task_success}, ORCA achieves a task success rate of 67.76\% in the cooperative scenario, outperforming all baselines.
Under the unseen malicious scenario, \ac{ORCA} maintains a success rate of 46.53\%, exceeding \ac{MADQN}, the best-performing \ac{MARL} method, by 8.12 percentage points. 
Regarding robustness, ORCA retains 68.7\% of its cooperative-scenario performance, whereas the learning-based baselines retain only about 58\%.
Across the five malicious-scenario runs, \emph{Learn} authorized 82 corrective executions. Of these, 11 successfully recovered tasks after a failed initial commitment. No corrective executions were authorized in the cooperative scenario.
This observation indicates that the reflective correction provides a bounded recovery path under failures, and the aggregate result primarily arises from the complete four-Agent decision process rather than repeated execution alone.
\subsubsection{System-level efficiency}
\
\newline
\indent 
\begin{figure}[tb]
    \centering        
    \begin{subfigure}{0.45\textwidth}
        \centering
        \includegraphics[width=\textwidth]{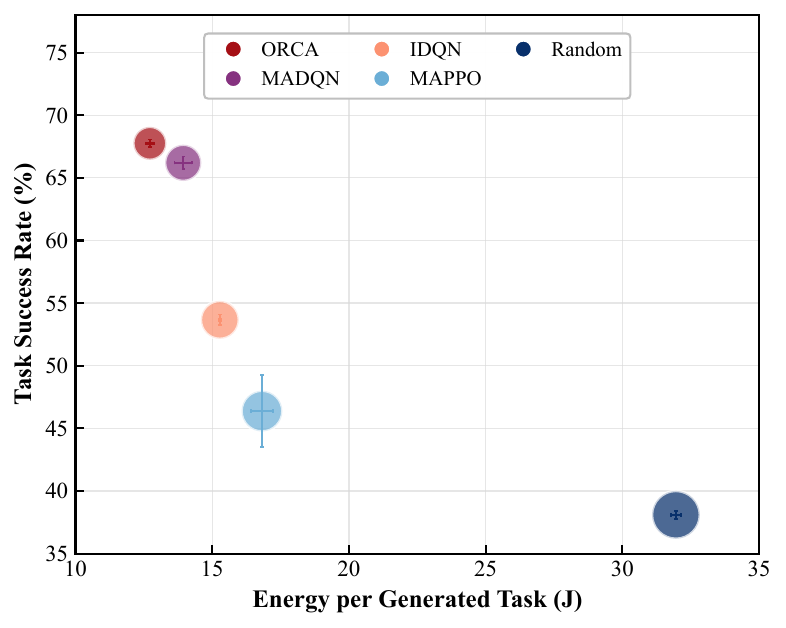}
        \caption{Comparison under the cooperative scenario.}
        \label{fig:efficiency_1}
    \end{subfigure}
    \begin{subfigure}{0.45\textwidth}
        \centering
        \includegraphics[width=\textwidth]{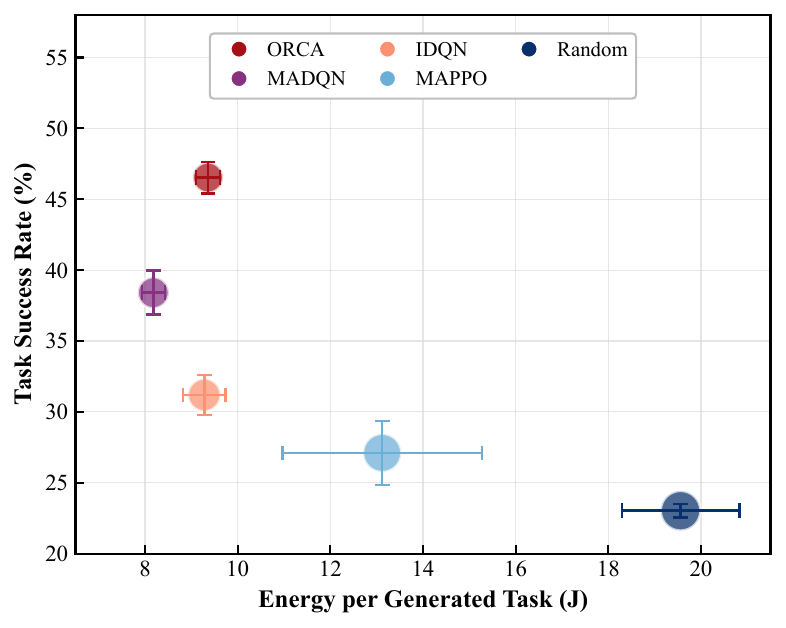}
        \caption{Comparison under the malicious scenario.}
        \label{fig:efficiency_2}
    \end{subfigure}
    \caption{Illustration of system-level efficiency of ORCA.} 
    \label{fig:efficiency}
\end{figure}
Fig.~\ref{fig:efficiency} compares system-level operating points in terms of task success rate and energy consumption. Bubble size represents the average \ac{E2E} latency.
In the cooperative scenario illustrated by Fig.~\ref{fig:efficiency_1}, \ac{ORCA} consumes 12.72 J per generated task with an average \ac{E2E} delay of 0.547 s, compared with 13.94 J and 0.748 s for \ac{MADQN}.
\ac{ORCA} raises success by 1.56 percentage points while reducing energy by 8.8\% and delay by 26.9\%.
This indicates that the selective task commitment of \ac{ORCA} reduces unnecessary remote transmissions and queuing overhead.
Fig.~\ref{fig:efficiency_2} compares the methods under the unseen malicious dropping scenario. \ac{ORCA} achieves a task success rate of 46.53\%, exceeding \ac{MADQN} by 8.12 percentage points. It also achieves lower average latency than \ac{MADQN}.
The additional energy overhead of \ac{ORCA} relative to \ac{MADQN} comes from reflective correction after collaboration failures and the subsequent execution of the affected tasks. 
Overall, \ac{ORCA} trades a moderate increase in energy consumption for higher task completion reliability and faster response under malicious conditions.

\subsubsection{Offloading selectivity}
\
\newline
\indent 
\begin{figure}[tb]
    \centering        
    \begin{subfigure}{0.45\textwidth}
        \centering
        \includegraphics[width=\textwidth]{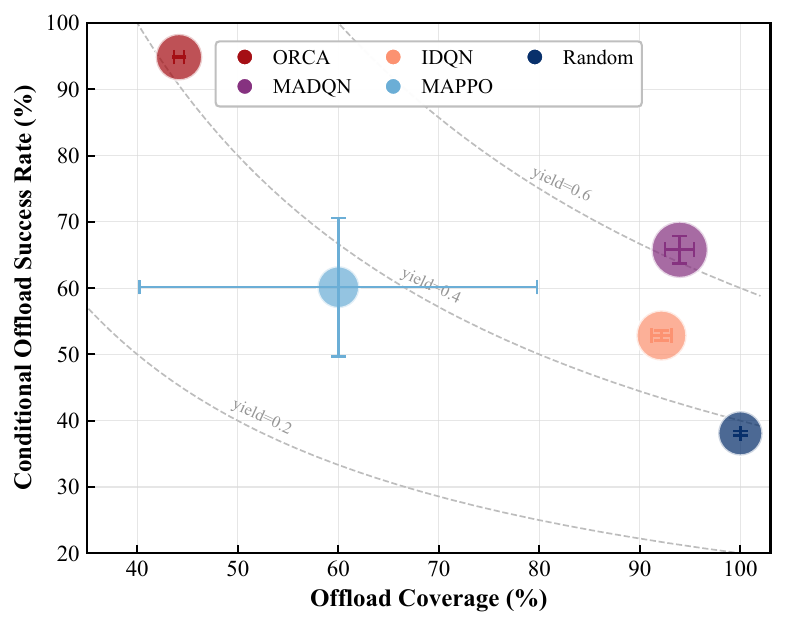}
        \caption{Comparison under the cooperative scenario.}
        \label{fig:offload_1}
    \end{subfigure}
    \begin{subfigure}{0.45\textwidth}
        \centering
        \includegraphics[width=\textwidth]{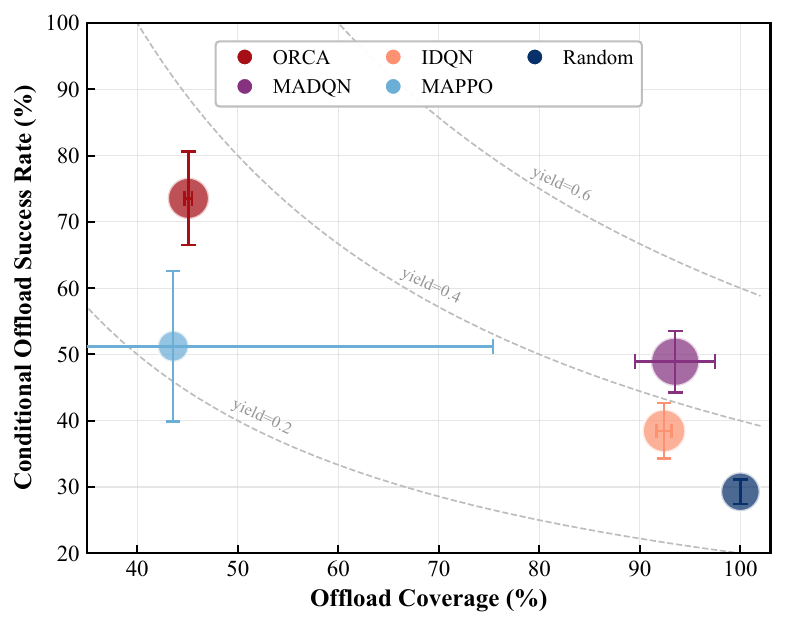}
        \caption{Comparison under the malicious scenario.}
        \label{fig:offload_2}
    \end{subfigure}
    \caption{Illustration of offloading selectivity and reliability of ORCA.} 
    \label{fig:offload}
\end{figure}
Fig.~\ref{fig:offload} separates offloading propensity from execution reliability. It reports offload coverage and conditional offload success rate. Their product gives the successful-offload yield, which is represented by bubble size.
Gray dashed lines in the figure represent curves of equal yield.
Compared with the high-frequency offloading behavior of the baseline methods, \ac{ORCA} exhibits a more selective offloading pattern. 
In the cooperative scenario shown in Fig.~\ref{fig:offload_1}, \ac{MADQN} achieves an offload coverage of 93.96\%, resulting in a relatively high remote-execution yield, but its conditional offload success rate is only 65.79\%. 
In contrast, \ac{ORCA} selectively offloads only 44.16\% of the tasks while increasing the conditional offload success rate to 94.84\%. 
More notably, in the unreliable scenario shown in Fig.~\ref{fig:offload_2}, fixed policies such as \ac{MADQN} experience a decline in conditional reliability to 48.90\%. 
By comparison, \ac{ORCA} maintains a conditional offload success rate of 73.54\% at an offload coverage of approximately 45\%. 
These results show that \ac{ORCA} is able to update neighbor-specific collaboration evidence and avoid repeated commitments to unreliable satellites, thereby maintaining higher offloading reliability in uncertain environments.

\section{Conclusion}
\label{sec:conclusion}
This paper introduces \ac{ORCA}, a four-Agent architecture for autonomous onboard task offloading in LEO satellite networks. 
By separating task interpretation, candidate reasoning, physical commitment, and outcome evaluation into a \emph{Perceive}$\rightarrow$\emph{Reason}$\rightarrow$\emph{Act}$\rightarrow$\emph{Learn} loop, \ac{ORCA} transforms offloading from a one-shot action into an outcome-grounded decision process with bounded reflective correction. 
In the cooperative scenario, \ac{ORCA} achieves a 67.76\% task-success rate while significantly reducing energy consumption and \ac{E2E} delay. 
Under the unseen malicious node-drop setting, \ac{ORCA} achieves the highest task-success rate of 46.53\%. It retains 68.7\% of its cooperative-scenario performance.
The offloading analysis further showed that \ac{ORCA} does not pursue remote execution frequency alone. 
Instead, it combines selective offloading with local execution and corrective reasoning, thereby maintaining a higher conditional reliability when previously cooperative neighbors become unreliable.

The current evaluation is based on a ground simulation of the spaceborne operating environment, and the Qwen inference delay and energy costs are configured rather than measured on flight hardware. 
Future work will conduct hardware-in-the-loop experiments with flight-representative satellite payload boards. These experiments will combine orbital and link emulation with measurements of actual \ac{LLM} inference latency, energy consumption, and thermal behavior.
Moreover, as spaceborne computing platforms continue to improve, more capable onboard processors are expected to support higher-performance LLMs, allowing autonomous Agents to reason over richer multimodal observations, coordinate more complex computation tasks and execute increasingly sophisticated missions directly in orbit.

\section*{Acknowledgement}
This work is supported by National Natural Science Foundation of China under Grant 62401070.

\bibliographystyle{IEEEtran}
\bibliography{reference}

\end{document}